# The Kep-Cont Mission: Continuing the observation of high-amplitude variable stars in the Kepler field of view
## Proposal for the Kepler white paper call

L. Molnár[1], R. Szabó[1], K. Kolenberg[2,3], T. Borkovits[4,5], V. Antoci[6,] K. Vida[1], C. C. Ngeow[7], J. A. Guzik[8], E. Plachy[1], B. Castanheira[9]

[1]*Konkoly Observatory, Research Center for Astronomy and Earth Sciences of the Hungarian Academy of Sciences, H-1121 Konkoly Thege Miklós út 15-17, Budapest, Hungary*
[2]*Harvard-Smithsonian Center for Astrophysics, 60 Garden Street, Cambridge, MA 02138, USA*
[3]*Instituut voor Sterrenkunde, University of Leuven, Celestijnenlaan, 200D, B 3001, Heverlee, Belgium*
[4]*Baja Astronomical Observatory, H-6500 Baja, Szegedi út, Kt. 766, Hungary*
[5]*ELTE Gothard-Lendület Research Group, H-9700 Szombathely, Szent Imre herceg út 112, Hungary*
[6]*Stellar Astrophysics Centre, Aarhus University, Ny Munkegade 120, DK-8000 Aarhus C, Denmark*
[7] *Graduate Institute of Astronomy, National Central University, Jhongli 32001, Taiwan*
[8]*Los Alamos National Laboratory, Los Alamos, NM  87545 USA*
[9]*University of Texas at Austin, Austin, TX 78712, USA*

**Abstract**

As a response to the Kepler white paper call, we propose to **keep Kepler pointing to its current field of view** and continue observing thousands of large amplitude variables (Cepheid, RR Lyrae and delta Scuti stars among others) with high cadence in the Kep-Cont Mission. The degraded pointing stability will still allow observation of these stars with reasonable (better than millimag) precision. The Kep-Cont mission will allow studying the nonradial modes in Blazhko-modulated and first overtone RR Lyrae stars and will give a better view on the period jitter of the only Kepler Cepheid in the field. With continued continuous observation of the Kepler RR Lyrae sample we may get closer to the origin of the century-old Blazhko problem. Longer time-span may also uncover new dynamical effects like apsidal motion in eclipsing binaries. **A continued mission will have the advantage of providing unprecedented, many-years-long homogeneous and continuous photometric data of the same targets.** We investigate the pragmatic details of such a mission and find a number of advantages, especially the minimal need of reprogramming of the flight software. Another undeniable advantage of the current field of view is the completed, ongoing and planned ground-based follow-up observations and allocated telescope times focusing on the current field. **We emphasize that while we propose this continuation as an independent mission, we can easily share slots with e.g. planetary mission with a strong belief that both (or more) communities can still benefit from Kepler's current capabilities.**

1. Introduction

Kepler ceased to observe its field-of-view in May 2013 after the second reaction wheel out of four showed elevated friction levels. NASA engineers tried to get one of the wheels back online, but after months of preparation and testing, it became clear that the the wheels are permanently compromised and the pointing accuracy required for the extraordinary photometric precision cannot be maintained. Indeed, the Kepler space telescope (Borucki et al. 2010) opened a new window to exoplanetary systems (Batalha et al 2013), as well as allowed an unprecedented view on stars themselves (Gilliland et al. 2010, Bedding et al. 2011). After the unsuccessful recovery attempts a white paper call[1] was announced to propose new missions exploiting the capabilities of Kepler handicapped by a reduced pointing accuracy. This work is an answer from the variable star community for the call.

## 2. Science goals

### 2.1. RR Lyrae stars

Relatively small space telescopes already delivered new results on high-amplitude pulsating stars. The Canadian MOST (Microvariability and Oscillations of STars) telescope for example monitored a double-mode RR Lyrae star, AQ Leo (Gruberbauer et al. 2007) establishing the existence of a new periodicity most probably originating from the presence of nonradial modes. CoRoT (Convection, ROtation and planetary Transit) continued the exploration and established new standards in the observations of amplitude modulated stars (Chadid et al. 2010, Guggenberger et al. 2011). The Blazhko effect is one of the longest standing mysteries in stellar pulsation theory. The enigmatic amplitude and phase modulation of roughly half of the RR Lyrae stars pulsating in the fundamental radial mode was discovered more than a century ago (Blazhko 1907, Shapley 1916).

The unprecedented Kepler Mission crowned the monitoring of Blazhko stars and non-Blazhko stars (Benkő et al. 2010, Nemec et al. 2011), unveiling new dynamical phenomena, like the period doubling (Kolenberg et al. 2010, Szabó et al. 2010), unexpected triple-mode state (Molnár et al. 2012) and even low-dimensional chaos (Plachy et al. 2013). These new discoveries induced theoretical studies as well (Szabó et al. 2010, Kolláth et al. 2011, Smolec & Moskalik 2012) leading to a new explanation of the Blazhko effect by a radial resonance between the fundamental mode and a high overtone (Buchler & Kolláth 2011). With the Kep-Cont mission the validity of these theories could be checked. In order to do that we plan to search for the presence of period doubling in the modulated RR Lyrae light curves which is caused by the very same resonance (Szabó et al. 2010, Kolláth et al. 2011), and analyze their characteristics. We also plan to run full hydrodynamical calculations to reproduce the full spectrum of dynamical behaviors and compare them with the Kepler and Kep-Cont observations.

Continued observations of RR Lyrae stars would help to study long-period Blazhko cycles. Also, it would allow investigating of the dynamical nature of the modulation thoroughly. In many cases we see irregular and non-repetitive modulation cycles, and we suspect that at least some of the modulated RR Lyrae stars display chaotic dynamics in the Blazhko-effect (cf. Plachy et al. 2013 and Buchler & Kolláth 2011). However, with the current amount of Kepler data is barely enough to detect low-dimensional chaos even for the shortest modulation periods (20-30 days). The majority of the Blazhko sample would enormously benefit from 3-4 years of more observations. Especially, the four-year cycle in the Blazhko period of RR Lyrae itself (Detre & Szeidl 1973) could be tested, since so far hardly one cycle has been covered with Kepler short cadence data. We note here that for the Blazhko cycles where the fine details of the light curve are less important, the degraded pointing accuracy would still ensure high enough precision to analyze them in detail. Further advantage is that the chaos analysis is only sensitive to the number of cycles, but not to the long gap between the primary and extended missions.

Another possibility with Kep-Cont would be to follow the temporary nature of the presence of the radial overtones and other extra frequencies (possibly non-radial modes) in RR Lyrae stars (Molnár et a. 2012, Benkő et al. 2010, Moskalik et al. 2013), this way we might get closer to the understanding of their excitation mechanism and coupling to other modes. In this respect it would be interesting to follow the more regularly pulsating non-Blazhko RR Lyrae star light curves (Nemec et al. 2011) to check their long-term light curve stability. Any deviation from clock-work precision might indicate stellar evolutionary effects or direct information on our imperfect understanding of physics in stellar interior, e.g. turbulence and convection.

---

[1] http://keplerscience.arc.nasa.gov/docs/Kepler-2wheels-call-1.pdf

More than a dozen new RR Lyrae stars (Szabó et al. 2013 in prep.) were discovered by Kepler serendipitously while analyzing other targets, by various Guest Observer programs and the PlanetHunters (Fischer et al. 2012). Some of them pulsate in the first overtone pulsation mode (RRc), many of them Blazhko-modulated. We plan to analyze these as well following the same lines as above.

### *2.2. Cepheid*

The nature of period jitter found in the case of V1154 Cygni (Derekas et al. 2012), has not been seen in other fundamental-mode Cepheids. With longer data any regularities or periodic cycles in the variation could be revealed that would in turn constrain the possible mechanisms explaining the unusual phenomenon.

### *2.3. Eclipsing binaries and triple systems*

In the Kepler field about a dozen unique and unprecedented triple (or multiple) systems were found where extra eclipses of further, more distant components were also detected. (The first two of such systems were KOI-126, Carter et al. 2011, and Trinity, Derekas et al. 2011). These systems, and some tens of others exhibit short period, low amplitude ETVs (Eclipse Timing Variations) coming from gravitational interactions (perturbations) between the stars, which makes it possible to determine stellar masses and 3D orbits from ETV curve alone (Steffen et al. 2011, Rappaport et al. 2013). It is a similar, but not identical method than the TTV analysis applied in multiple exoplanetary systems. These discoveries offer an opportunity to follow-up these systems to see the variations in their orbital elements. That can be done with the reduced precision for almost all systems: the eclipses reach 1% in depth in approximately 90% of the targets. Two examples are plotted in Figure 1.

There are also binary systems, where parabolic eclipse timing variations (ETV) were observed (Tran et al. 2013, Conroy et al. 2013). Such variations may have different origins, such as (i) light-time effect caused by a tertiary component, (ii) magnetic cycles (Borkovits et al. 2005), (iii) mass exchange, or mass-loss and (iv) tidal dissipation (Borkovits et al. 2011). To discriminate between them, the quantitative analysis of ETVs requires longer time-span observations.

Indeed, one of the main breakthroughs of Kepler observations on eclipsing binaries does not come only from the unprecedentedly accurate light curve observations, but also from the long-term continuous cycle-to-cycle eclipse observations that makes it possible to determine timing data with high accuracy, and with an almost 100% cycle-coverage. After four years of observations apsidal motion and longer period dynamical variations (due to perturbations) are just about to be detectable in several Kepler eclipsing binary systems. We propose to start looking for these effects if Kep-Cont is approved.

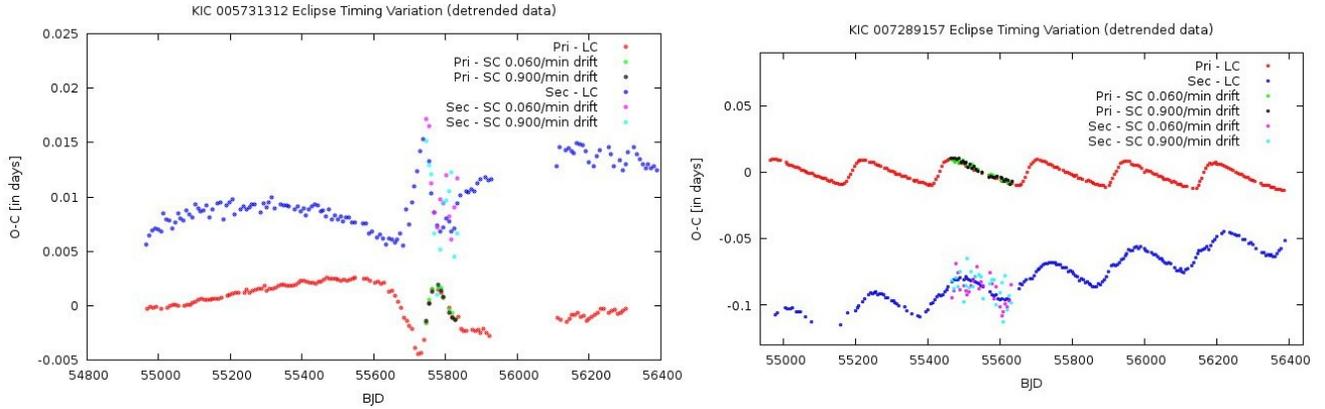

*Figure 1. Effects of the simulated drift noise on the ETV calculations. A single quarter of short cadence data was modified with the noise for both stars and the resulting fits were compared to the ETV calculations based on the long-cadence data. Primary eclipses can be timed with almost the same accuracy. Shallow (1-2% deep) secondary eclipses are affected but the variations are still detectable. The data is based on the current pipeline: additional post-processing will increase the accuracy for the shallow eclipses too.*

### 2.4. Delta Scuti, gamma Doradus stars and other A-F-type pulsators

The range of A-F spectral classes, just above the Main Sequence, contains an interesting mixture of variables: delta Scuti, gamma Doradus and hybrid stars, rho Puppis, SX Phe Am, Ap and roAp-type stars, all of which are of great interest. The reason why these stars are so important is because they occupy the region in the HRD where on the one hand the transition from deep and effective to shallow convective envelopes takes place and on the other hand the rotational velocity is observed to increase from an average of 10 to 100 km/s (Royer 2009). The two processes, i.e. rotation and convection are, of course, not independent of each other and have a great impact on mixing processes, transport of angular momentum, magnetic fields, activity, pulsation, and so on.

Although the degraded photometric accuracy will not allow for further investigation of the apparently non-variable population of stars and the purity of the instability strips, continued observations will be still valuable. Variations in mode amplitudes may be important tools for detecting mode interactions and helping the mode identification (Balona et al. 2012). Slow, persistent mode amplitude changes in HD 178875 were explained by changes in the stellar structure (Murphy et al. 2012). This star provides a very rare opportunity to detect stellar evolutionary changes over human time-scales, hence worthy of extended follow-up. There may be other stars where the 4 years of the primary mission was insufficient to detect such changes but extended coverage may uncover similar evolutionary effects.

Flares and flare-like signatures were observed in a wide range of spectral classes. Unexpectedly, early-F and A-type stars were found to display flares too (Balona 2012). These stars are thought to lack a corona and the flares may have different origin. Further observations can help to shed light on this topic, although the amplitudes of flares in A-type stars are close to the simulated drift noise.

Kepler observations (e.g. HD 187547, Antoci et al. 2011; Uytterhoeven et al. 2011) demonstrated that we still need to better understand the excitation mechanisms operating in A and F type stars. Monitoring the stars showing the largest amplitudes of pulsations will allow us to study their stability.

### 2.5. Compact pulsators

Uninterrupted, high-precision observations of subdwarf B stars and pulsating white dwarfs proved to be extremely valuable in the understanding of these stars. The rich frequency spectra allows to detect the correct g-mode spacing and to carry out the mode identification (Reed et al. 2011). Fundamental parameters (mass, luminosity, radius, atmospheric parameters, composition of the envelope and the core) may be determined from asteroseismic modeling (Charpinet et al. 2011).

Compact pulsators will benefit extraordinarily from the extended mission. These stars have short pulsation periods that are only detectable with short cadence. In the early years of the mission, every candidate was observed for a short period (a month or a quarter) only, to select the best candidates for later follow-up. However, compact pulsators are known to show significant changes in their oscillation spectra and mode amplitudes (e.g. Kilkenny et al. 1999, Montgomery et al. 2010). With the continuous, short-cadence observations of Kep-Cont, such drastic changes will be observable while they are happening.

Due to their intrinsic faintness, only a few pulsating white dwarfs are known in the Kepler field. Most were discovered by targeted surveys during the mission and were included in the target list late: some were not observed before sudden the end of the primary mission. Because of their relatively simple internal structure, pulsating white dwarfs are the only class of stars accessible for ensemble asteroseismology studies (Castanheira & Kepler 2009). Hence the continued mission, with all stars included, would advance the understanding of white dwarfs too.

### 2.6. Other types of variable stars

The research of stellar activity would also greatly benefit from the continued observation of the current Kepler-field. This way we could extend the time base of the data sets, and study the evolution of active stars. The current length of the observations enables us to detect the shortest existing activity cycles, and even these detection would need further observation to be confirmed. Evolution of starspots is closely related to activity cycles. Massive stars, on the hot (beta Cephei, SPB stars) and the cool ends of the HRD (Miras and semiregular stars) will also benefit from more pulsation cycles that are observed.

### 2.7. Stellar population studies

There are four open clusters in the Kepler field, spanning a large range of ages (between 0.7 and 8.3 Gyr) and different metallicities. The homogeneous sample of cluster members allows for population studies of several types of stars. Two clusters were observed with superstamps, single large apertures covering the clusters, but from the other two, only selected stars were measured. Although the centers of those clusters are crowded, we believe that more homogeneous comparison can be achieved by placing superstamps on all clusters.

## 3. Observing strategy

Length of the proposed mission: 2-4 years. Kepler will continue the observations of the original field to advance our understanding of the stars and planets there. New targets may also be allocated.

### 3.1. Pointing

The original Kepler field can be maintained with 1-day-long pointing attitudes and daily thruster firings with an expected drift/roll angle of ~2 arcminutes. That translates to a few pixels shift in the focal plane, 3-4px at the edges of the field-of-view. However, the spacecraft attitude has to be maintained with respect to the Sun to minimize solar torque: the boresight will rotate with ~1 degree at every pointing tweak. The interval between pointing adjustments can be lengthened up to 4 days, if necessary, but based on the Appendix for the Call, drift rates will increase farther from the balance point.

Fuel on-board may last to 2-4 years, depending on the pointing requirements. Such conditions make the extended Kepler mission feasible. The minimal mission duration would be 1 year.

### 3.2. Expected photometric accuracy

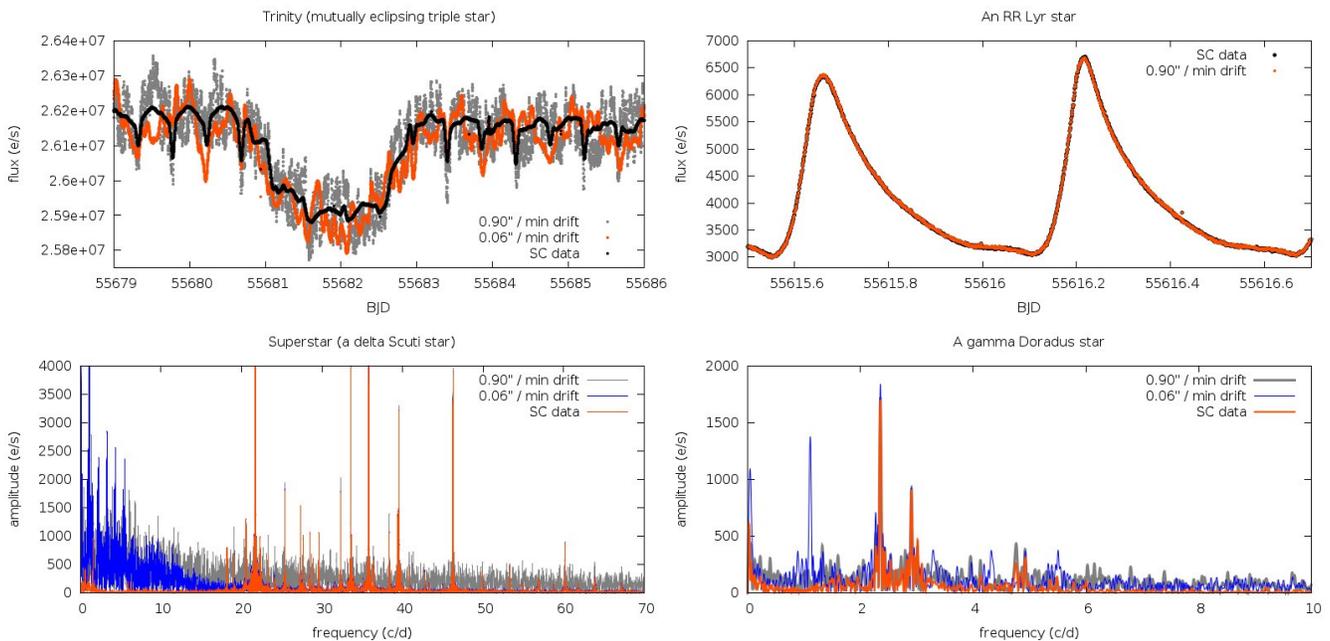

*Figure 2. Effects of the simulated drift on various variables. RR Lyrae stars (top right) are largely unaffected thanks to their high intrinsic amplitudes. Slower and faster drift rates compromise different variable stars: the secondary eclipses in Trinity (HD 181068), a mutually eclipsing triple star (top left), are obscured by the slow drift. The typical frequency regimes of delta Scuti (bottom left) and gamma Doradus pulsations (bottom right) are also affected differently (note the strong false peak in the gamma Dor frequency spectrum).*

A detailed simulation of the expected photometric accuracy was carried out within the KASC[2]. Global pixel-to-pixel variations, intrapixel variations and sensitivity drops due to pixel-channels based on the Kepler Instrument Handbook were incorporated into the Kepler photometry simulator (De Ridder, Arentoft & Kjeldsen, 2006). According to the results, the drift introduces variations on the levels of ± 4000 ppm with characteristics that highly depend on the actual drift rate. This drift noise will probably be too high to detect the intricate stellar oscillation signals but other stars with higher amplitudes and/or variation time scales that differ sufficiently from the characteristics of the noise may still be observable. Large-amplitude variability, either slow or fast is mostly unaffected by the drift except for the finest details in the light curves. Depending on the actual drift rate, low-amplitude, high-frequency variations (e.g. pulsations in white dwarfs and delta Scuti variables) may be also preserved.

Photometric accuracy will be further affected by the roll of the field-of-view as stars may move onto different modules over time. However, these estimates are still very preliminary and the actual noise levels will strongly depend on the telescope performance and post-processing capabilities. We expect that the noise can limited to ~1000 ppm or lower, especially in the middle of the field-of-view. The best ground-based observations may reach similar precision, but only for a limited number of targets and of course nor continuously over the entire year, emphasizing the importance of space-based observatories like Kepler.

### 3.3. *Integration times and masks*

We propose to observe most (if not all) targets in short cadence. Slow drift rates affect the photometric precision in the high-frequency range – above the long cadence Nyquist-frequency – the least. Given the limitations of the data recorder, increasing the number of SC targets will limit the overall number of available pixels. Rotation of the field-of-view will require either long arc-shaped masks or the ability to shift/reposition the masks when needed. We understand that the latter scenario requires changes to the flight software but strongly favor this solution in order to maximize the number of simultaneous targets.

Two open clusters out of four were observed with so-called superstamps, extended masks that covered the entire clusters. We propose to continue the observations of all four clusters with superstamps and long cadence sampling.

### 3.4. *Number of targets, ground-based follow-up*

Target selection for the continued observations will be a relatively straightforward exercise that may be refined once the photometric accuracy is determined. The original KASC target list contained approximately 7000 targets. A large part of those were solar-like stars and red giants that are probably not feasible any more. On the other hand, several new variables were identified among the ~170 000 stars Kepler has observed. Still, the number of targets that are both feasible and require extended coverage will probably fit into a short-cadence-only mission, augmented with the four superstamps. If the photometric accuracy will exceed the current expectations, the number of short- (and possibly long-)cadence targets can be adjusted to incorporate other targets (solar-like oscillators, exoplanet hosts) as well.

---

[2]Kepler Asteroseismic Science Consortium, http://astro.phys.au.dk/KASC/

Another advantage of the original field is the extended, still ongoing ground-based follow-up. Multicolor photometry and spectra are important additions to the white-light observations of Kepler. Data have been or are being collected by several observatories and surveys, limiting the need for follow-up during the Kep-Cont mission.

*3.5.     Synergies with the possible planetary and stellar mission(s)*

While we propose Kep-Cont as an independent mission, this program can easily share slots with other proposals, e.g. a planetary mission or follow-up on solar-like oscillations targeting the same field. Slots can be shared, slots of dropped targets can be reused, and a common strategy can be followed concerning the observations (pointing control, apertures, cadence, etc.). A prioritized target list that we can provide with Kep-Cont - if selected - would make it easy to merge with missions with other scopes.

**4.     Conclusions**

The observations collected by the original Kepler mission, spanning up to 4 years almost continuously for ~170 000 stars, is so far unprecedented, and will continue to provide many discoveries in the following years. However, we believe that Kepler can do even better. The telescope is still able to collect photometric data almost continuously although with lower accuracy. In this white paper we argue that even if a notable part of the targets (solar-like oscillators and the smallest exoplanets) may not be observable any more, **there is still a wide variety interesting objects and in the original Kepler field-of-view that justifies the Kep-Cont extension of the original mission.**

Almost all classes of pulsating variables, RR Lyrae stars, pulsating stars along the Main Sequence and compact pulsators, hot and cool massive stars can be followed and longer time-spans may help to answer several questions about them. Eclipsing binaries, triple systems will also greatly benefit from the extended coverage by detecting slower dynamical effects. Observations of all four open clusters with full superstamps (extended apertures covering entire clusters) instead of only a selection of stars will allow for more detailed population studies.

A major change with respect to the original mission would be the extensive use of short-cadence integrations: we propose to observe 10 times more stars with one-minute sampling than in the original mission. This will of course reduce the number of targets, but the actual allocation will have to be based on the performance of the telescope. Although we propose Kep-Cont as a standalone mission, our proposal is flexible, and **we are open to the synergies with other fields, such as planetary missions.**

The expertise gained in the primary mission by the scientific community, the minimal to modest efforts in target selection and flight software changes, the ongoing ground-based follow-up programs and the already solid science behind the mission make the Kep-Cont a worthy extension of the original mission.

**Acknowledgements:**
We gratefully acknowledge the following grants: the Lendület-2009 Young Researchers' Program of the Hungarian Academy of Sciences, the HUMAN MB08C 81013 grant of the MAG Zrt., the Hungarian OTKA grant K83790 and the KTIA URKUT_10-1-2011-0019 grant. We wish to thank organizers of the IAU Symposium 301 (Precision Asteroseismology: Celebrating the scientific opus of Wojtek Dziembowski) for allowing us to organize a splinter session on the future of Kepler. L. M., R. Sz., K. K., J. A. G., E. P. and V. A. acknowledge the IAU grants to the conference. R. Szabó acknowledges the János Bolyai Research Scholarship of the Hungarian Academy of Sciences. Work of L. Molnár leading to this research was supported by the European Union and the State of Hungary, co-financed by the European Social Fund in the framework of TÁMOP 4.2.4. A/2-11-1-2012-0001 'National Excellence Program'.